\documentclass[12pt]{iopart}

\usepackage{graphicx}

\begin{document}

\title{Jet quenching via jet collimation}

\author{J Casalderrey-Solana$^{1}$ , \underline{J G Milhano}$^{1,2}$ and  U Wiedemann$^{1}$}

\address{$^1$Physics Department, Theory Unit, CERN, CH-1211 Gen\`eve 23, Switzerland}

\address{$^{2}$ CENTRA, Departamento de F\'isica, Instituto Superior T\'ecnico,\\
Universidade T\'ecnica de Lisboa,
Av. Rovisco Pais 1, P-1049-001 Lisboa, Portugal}

\ead{guilherme.milhano@ist.utl.pt, jorge.casalderrey@cern.ch, urs.wiedemann@cern.ch}

\begin{abstract}
The strong modifications of dijet properties in heavy ion collisions measured by ATLAS and CMS provide important constraints on the dynamical mechanisms underlying jet quenching. In this work, we show that the transport of soft gluons away from the jet cone   --- jet collimation --- can account for the observed dijet asymmetry with values of $\hat{q}\, L$ that lie in the expected order of magnitude. Further, we show that the energy loss attained through this mechanism results in a very mild distortion of the azimuthal angle  dijet distribution.
\end{abstract}

The kinematical reach of the LHC allows for systematic studies involving fully reconstructed calorimetric jets to be carried out, and for the ensuing analyses to be significant beyond the experimental uncertainties inherent to the identification of jets in a fluctuating high multiplicity environment. 

The measurement of the dijet asymmetry $A_J= (E_{T_1}-E_{T_2})/(E_{T_1}+E_{T_2})$ in Pb-Pb collisions at $\sqrt{s}=2.76$ TeV by both ATLAS \cite{Aad:2010bu,atlasjets} and CMS \cite{Chatrchyan:2011sx,cmsjets} provides a clear example of the extent to which such full jet measurements can contribute to furthering  the understanding of the dynamical processes responsible for jet quenching. In the analyzed dijet event samples, the transverse energies of the jets are identified within cones of radius $R$ (0.4 for ATLAS, 0.5 for CMS). For the leading jet, it is $E_{T_1}> 100$ GeV (ATLAS), $120$ GeV (CMS), while for the recoiling jet --- found at an azimuthal separation $\Delta \Phi =|\Phi_1 -\Phi_2|>\pi/2$ (ATLAS) and $2\pi/3$ (CMS) --- it is $E_{T_2}> 25$ GeV (ATLAS), $50$ GeV (CMS).

The measured event asymmetry distribution shows qualitative features consistent with a substantial medium induced energy loss of the recoiling jet. In a nutshell, the fraction of energy lost from the recoiling jet cone is larger in heavy ion collisions than in the proton-proton case, and grows with increasing centrality of those collisions. 
Crucially, this effect is accompanied by a distortion of the dijet azimuthal distribution which is, at most, very mild.
 As such,  this measurement carries the hallmark of an underlying dynamical mechanism capable of carrying a substantial amount of energy away from the jet cone without any significant deflection of the jet direction.

In the 10\% most central events (for which the observed effect is stronger), the average amount of energy lost from the recoiling jet cone in addition to the proton-proton case falls within the bounds  \cite{us}
\begin{eqnarray}
\label{eq:enatlas}
10\, \mbox{GeV}  &<\Delta E < 21\, \mbox{GeV},  \qquad \mbox{(for the ATLAS data in \cite{Aad:2010bu})}\, , \\
\label{eq:encms}
8\, \mbox{GeV}  &<\Delta E < 18\, \mbox{GeV}, \qquad \mbox{(for the CMS data in \cite{Chatrchyan:2011sx})}\, .
\end{eqnarray}
Here, the lower bounds were obtained by assuming that all recoiling jets traverse the medium. The upper bounds resulted from taking only a fraction ($\sim 0.5$, the ratio of symmetric $A_J=0$ dijet events in Pb-Pb and proton-proton) as interacting with the medium.

\begin{center}
\begin{figure}[t]
\includegraphics[width=0.35\textwidth]{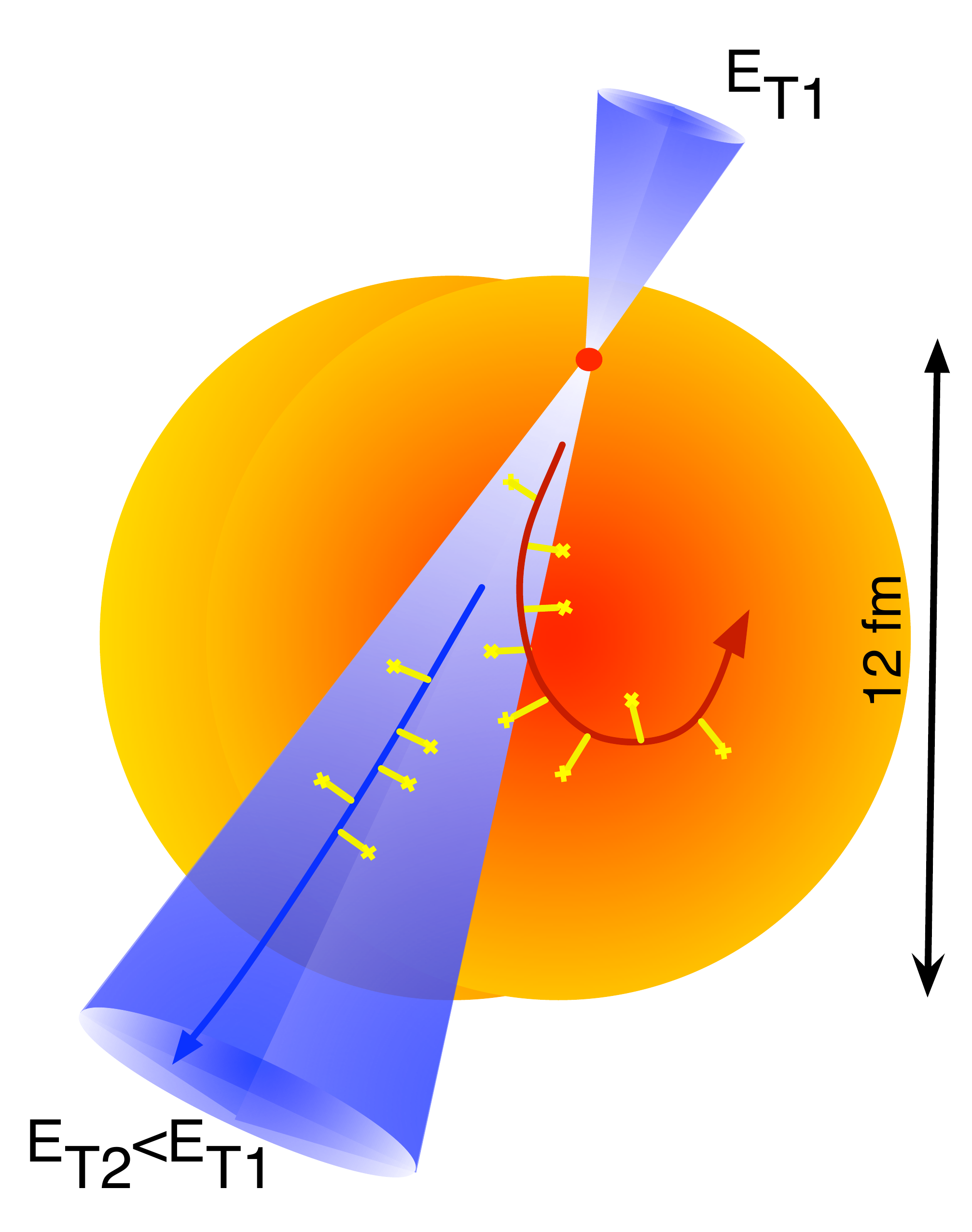}
\hspace{1.5cm}
\includegraphics[width=0.55\textwidth]{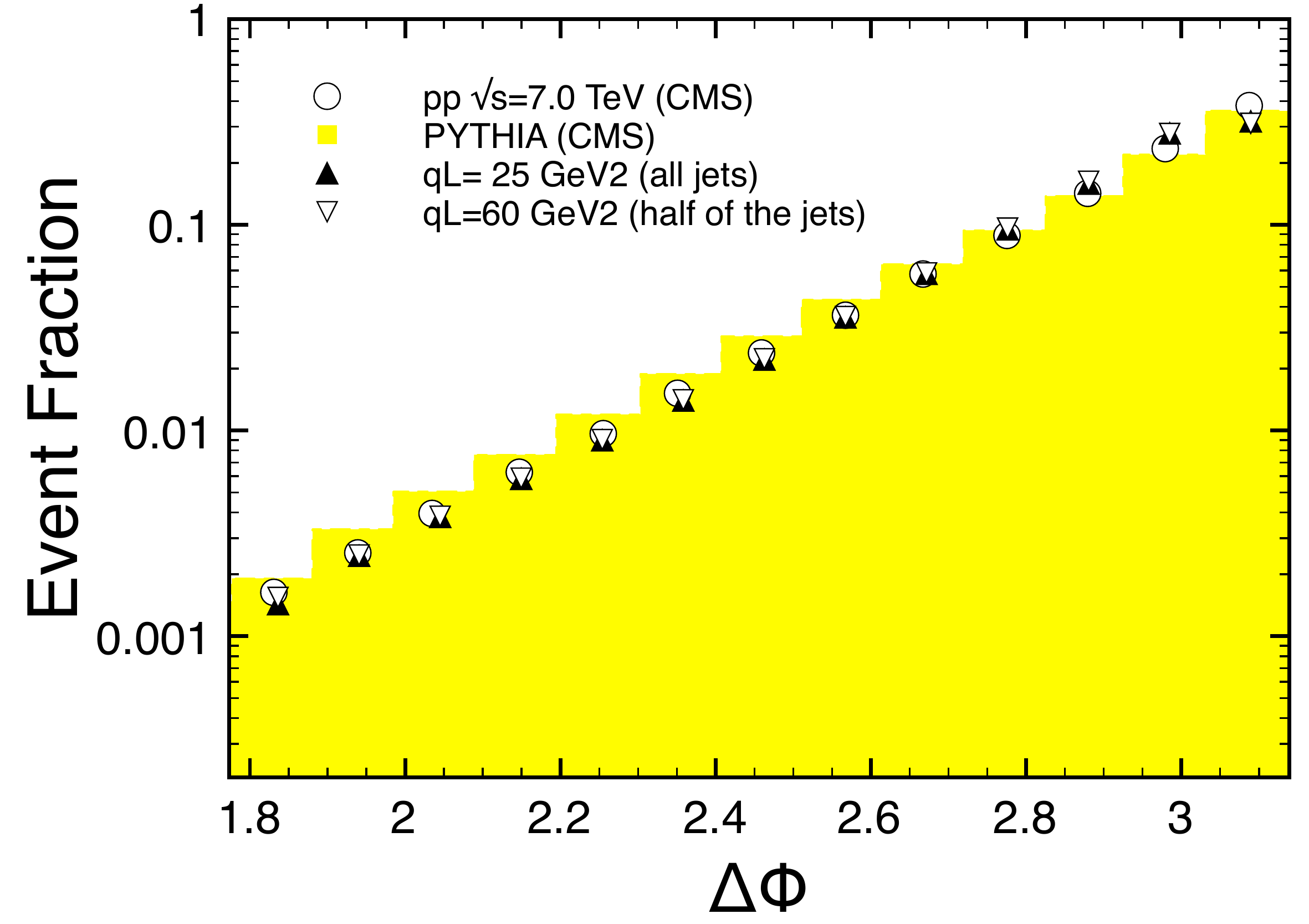}
\caption{\label{fig:}(left) Jet collimation: soft gluons emmited at small angles are transported outside the jet cone by their multiple random scatterings with medium components. (right) Azimuthal angle distribution of dijet events: the considered values of $\hat q L$ correspond to the cases in which either all recoiling jets are taken as interacting with the medium or only a fraction ($\sim$ 0.5) do.}
\end{figure}
\end{center}

A mechanism --- jet collimation --- that leads to both large energy degradation and no sizeable azimuthal displacement was proposed in \cite{us}. Soft gluons radiated at small angles result in 
a very small 
azimuthal displacement of the jet. These gluons, and in fact all partons in the parton shower, undergo Brownian motion during their passage through the medium and thus accumulate an average squared transverse momentum $\hat{q}$ per unit path length, $\langle k_T^2 \rangle \sim \hat{q} L\sim \hat{q} \tau$. Thus, in the presence of a medium, the average formation time $\tau\sim \omega/k_T^2$ for partons of energy $\omega$ is 
\begin{equation}
	\langle \tau \rangle \sim \sqrt{\frac{\omega}{\hat{q}}} 
\end{equation}
As a result, soft jet fragments are formed early and those with energy $\omega\leq \sqrt{\hat{q} L}$ will be completely decorrelated from the initial jet direction, see Fig.\,\ref{fig:} (left). In other words, the medium acts as a frequency collimator, efficiently trimming away the soft components of the jet by transporting them to large angles. This is in agreement with the observation by CMS  \cite{Chatrchyan:2011sx,cmsjets} that the energy lost from a jet cone is fully recovered at large angles and in the form of soft partons.

In \cite{us}, we showed that jet collimation can displace a sufficiently large fraction of the total jet energy outside the cone. It should be emphasized that jet collimation is effective even in the absence of any other medium induced effects. In other words, even in the extreme case in which the interaction with the medium does not result in any modification of the parton shower, jet collimation will result in the transport of a significant amount of energy away from the jet cone.

In this case, one can take the vacuum intra-jet multiplicity distribution
 $dD/dz$ 
 in the modified leading log approximation (known to be sufficiently accurate for small longitudinal momentum fractions $z\approx\omega/E_T$) to compute the energy fraction carried by partons with an energy fraction smaller than $z$
\begin{equation}
	\frac{E(z)}{E_T}=\int^\infty_{\log{1/z}} d\xi\, e^{-\xi} \frac{dD}{d\xi}\,.
	\label{eq:energy}
\end{equation}
If jet collimation is sole medium modification, then $\hat{q} L$ can be estimated by determining the highest $z$ to which the integration in eq.(\ref{eq:energy}) must be extended so as to account for the energy loss in eqs.~(\ref{eq:enatlas}, \ref{eq:encms}). Thus, since $\omega^2 = z^2\, E_T^2 \leq \hat q L$, we obtain that 

\begin{eqnarray}
\label{eq:atlasqhatl}
35 &\,    \le\, \hat q L\, \le 85 \,   {\rm GeV^2}\qquad {\rm (ATLAS\,  \cite{Aad:2010bu})}\, ,\\
\label{eq:cmsqhatl}
24 &\,   \le\, \hat q L\, \le 62 \,   {\rm GeV^2}\qquad {\rm (CMS\,   \cite{Chatrchyan:2011sx})}\, .
\end{eqnarray}

The inclusion of other medium induced effects, such as medium induced radiation, leads in general to a softening of the intra-jet multiplicity. 
While these other effects may  not be important for the description of the jet asymmetry data, they are crucial in  accounting for the suppression of single particle observables. 
The inclusion of these effects leads to the generic observation that the 
 same energy loss would be achieved with lower values of $\hat q L$. In this sense, the estimates eqs.~(\ref{eq:atlasqhatl}, \ref{eq:cmsqhatl}) provide upper bounds. 
 Since there is not yet a consensus on the precise underlying dynamics leading to this softening
 \cite{Armesto:2011ht}, in \cite{us}, we use one particular  simple model \cite{Borghini:2005em} as an estimate of this effect:
 \begin{eqnarray}
\label{eq:atlasqhatl}
30 &\,   \le\, \hat q L\, \le 60 \,   {\rm GeV^2}\qquad {\rm (ATLAS\,  \cite{Aad:2010bu})}\, ,\\
\label{eq:cmsqhatl}
18 &\,    \le\, \hat q L\, \le 40 \,    {\rm GeV^2}\qquad {\rm (CMS\,   \cite{Chatrchyan:2011sx})}\, .
\end{eqnarray}
 
These values of $\hat q L$ will necessary lead to a modification of the azimuthal dijet distribution. For jet collimation to be a valid candidate mechanism of jet quenching, these modifications need to be sufficiently small as not to contradict the very mild distortion observed experimentally.
To estimate the effect of the medium induced broadening,
we take the reference azimuthal distribution from the proton-proton case in \cite{Chatrchyan:2011sx,cmsjets} as corresponding to an unmodified jet that can be embedded in a heavy ion event. The effect of jet collimation on this distribution can be assessed by smearing the transverse momentum corresponding to each azimuthal angle $p_T = \langle E_{T_2} \rangle \,\sin(\pi-\Phi)$ (with $ \langle E_{T_2} \rangle$ the average energy carried by a recoiling jet in central PbPb collisions) with a gaussian weight of average squared transverse momentum $\hat q L$. 
Following the two extreme models we have considered, the smearing is performed in all jets or only on a fraction ($\sim 0.5$) of them. 
The displaced momenta can them be recast as azimuthal angles.
The result of this procedure is shown in Fig.\, \ref{fig:} (right). There it is clear that even for the highest $\hat q L$ values consistent with CMS data, the distortion of the azimuthal distribution due to the transport of soft partons outside of the jet cone is very mild and not significant beyond the differences between proton-proton data (open circles) and its PYTHIA generated counterpart (solid histogram).

This estimate also shows the difficulty in determining the medium induced broadening of high energy jets, since even in vacuum there is a significant momentum imbalance of the dijet pair ($\left<k^2_T\right> \sim \mathcal{O} (100\,  {\rm GeV^2})$ for CMS jets). Thus, even for the largest medium induced momentum broadening we have considered, the most salient change in the dijet azimuthal distribution is a reduction of the yield at $\Delta \Phi\sim \pi$. The experimental quantification of this effect demands an accurate determination of this azimuthal distribution. However, such a measurement is complicated by the presence of the large hadronic background for jet reconstruction in lead-lead collisions which also broadens the dijet azimuthal distribution. While a precise determination of this effect requires a full study along the lines of  \cite{Cacciari:2011tm}, it is clear from the studies of PYTHIA jets into  heavy ion effects 
performed in  \cite{Chatrchyan:2011sx,cmsjets} that the effect of background particles is comparable to the one of momentum broadening.

In conclusion, the jet collimation mechanism proposed in \cite{us} is fully consistent with all available experimental data. In particular, it can account for the necessary energy loss from the jet cone without, as has been shown here, leading to an observable distortion of the azimuthal angle dijet event distribution. 
As such, the transport of soft jet components ought to be included in any formulation of jet quenching. A detailed qualitative assessment of jet collimation should proceed via its implementation in existing in medium Monte Carlo codes.

\vspace*{2mm}
\noindent\textbf{Acknowledgements} JGM acknowledges the support of Funda\c c\~ao para a Ci\^encia e a Tecnologia (Portugal) under project CERN/FP/116379/2010.

\end{document}